INVESTIGATION OF THE PROPORTIONAL DISCHARGE MECHANISM IN NONELECTRONEGATIVE GASES

*B.M.Ovchinnikov, V.V.Parusov, Yu.B.Ovchinnikov*

Institute for Nuclear Research of the Russian
Academy of Sciences, 60-th October
Anniversary prospect 7a, Moscow 117312, Russia,

Phone: (095) 3340194, (0967)517275,
FAX: (0967)517275,
E-mail: <ovchin@al20.inr.troitsk.ru>

Abstract

A shape of the signals in proportional chamber with preamplification gap and electron component extraction from proportional discharge has been investigated. It was shown for Penning mixtures that in proportional discharge besides the primary avalanches there is also the tail of secondary avalanches caused by metastable and nonmetastable Penning effects. For different mixtures of gases the duration of the signals caused by secondary avalanches are equal to $\sim n \cdot 10$ μs ($n \cong 1 \div 10$), the total charge of secondary avalanches is comparable or larger than in primary avalanches.

The generally accepted picture for signal formation in proportional counters and chambers is the following. The ionization electrons which drift in the electric field to the anode wire, are multiplied in the primary avalanches in high electric field near the anode and are collected at the anode. The production of exited states of gas atoms and molecules takes place in the process of the avalanches developing. For the mixture of noble gas X with the gas M which has the ionization potential less than the energy of exited metastable states of NG -atoms or less than the energy of photons, emitted by exited NG-atoms and molecules, (the Penning mixtures) the secondary avalanches are produced owing to collisions of long living metastable NG-atoms $X_m$ with atoms or molecules of the gas M:

$$e^- + X \rightarrow X^*_m + e^- \quad (1)$$

$$X^*_m + M \rightarrow X + M^+ + e^-, \quad (2)$$

and owing to photons emitted by exited NG atoms and molecules:

$$X^* + X \rightarrow 2X + h\nu \quad (3)$$

$$X_2^* \rightarrow 2X + h\nu \quad (4)$$

$$h\nu + M \rightarrow M^+ + e^-. \quad (5)$$

These secondary processes are described in works with proportional counters and chambers [1–5]. But the time structure of these processes was not measured because the shape of proportional signals in ordinary counters and chambers is mainly determined by slow motion of the positive ions from the anode, and the slow secondary signals can not be extracted from the positive ions signals in this case.

In this work the time structure of the electron component of the proportional signals is measured.

## I. Proportional chamber with extraction of the electron component of the signal

The proportional chamber with preamplification gap and with extraction of the electron component of the signal [6 – 8] was used in the measurements (Fig.1). The chamber containes the cathode 1 with an $\alpha(\beta)$-source 2 on it; the drift gap with rings 3 which provide a uniform electric field; the preamplification gap between cathode 4 and anode 5; the transfer gap 5-6 and the detecting induction gap 6-7. The cathode 1 has a 4 mm – diameter collimating opening, behind which a $^{239}Pu$ $\alpha$–source with a ~ $10^5$ $\alpha \cdot s^{-1}$ flux rate or a $^{63}$ Ni $\beta$ - source with a ~ $10^4$ $\beta \cdot s^{-1}$ flux rate is installed at a distance of 10 mm. The multiwire proportional gap (MWPG) used for electron multiplication contains the cathode 4 wound with a 2 mm pitch of a beryllium bronze 0.1 mm diameter wire. The anode 5 is winded of a 20 $\mu$m diameter W + Au wire with a 2 mm pitch. The MWPG is equal to 2 mm. In projection, the anode wires are parallel to cathode wires and exactly in the middle between them. The transfer gap is equal to 9 mm. The cathode 6 of the detecting gap is winded of a 0.1 mm diameter wire with a 1 mm pitch. The width of the gap 6-7 is 2 mm. The distribution of voltage between the electrodes of the chamber is determined by resistor divider shown in Fig. 1.

The electric field distribution in the chamber look like to that in the widely used chambers [7].

The ionization electrons from $\alpha(\beta)$ – particle track drift in the proportional gap and are multiplied at the anode wires 5. The electron charge in transfer gap 5-6 arises owing to electrostatic and photon mechanisms [7]. From transfer gap the electrons drift into the detecting gap 6-7. The electric field intensity in this gap is insufficient for evolving avalanches and the gap records the electrons without multiplication, i.e. in the induction operation mode of the ionization chamber.

The drift velocity of electrons in gap 6-7 is very high (~$10^6$ $cm \cdot s^{-1}$) for utilized gas mixtures and for electric field intensities in the gap (> 1 $kV \cdot cm^{-1}$). Owing to this the time resolution of the chamber is equal to ~ 0.2 cm / ($10^6$ $cm \cdot s^{-1}$) = $2 \cdot 10^{-7}$ s. Thus, the time structure of proportional discharge is measured with accuracy ~ $2 \cdot 10^{-7}$s by chamber shown in Fig. 1.

The signals from the anode of the detecting gap are fed to the charge – sensitive amplifier BUS2-96 that has the differentiating constant equal to 1 ms and after amplification the signals are detected by memory oscilloscope C 9-8.

During the measurements the chamber was being filled with different gases (see Table 1). The used gases were purified from impurities up to very low level. The oxygen was removed from all gases by an adsorbent $Ni/SiO_2$ up to a level of about $10^{-10}$ rel. vol. parts [9]. The gases He, Ne were purified of all impurities by fine filtration at a temperature of 78 K [10]. The Ar and Xe were purified of all chemical active impurities up to a level of ~ $10^{-9}$ rel. vol. parts by Ca at high temperatures (500-700$^o$ C) [11]. The Ar was also purified of all impurities up to a level of ~ $10^{-9}$ rel. vol. parts by synthetic zeolites NaA-6 at a temperature of 90 K [12].

The impurities $CH_4$, $C_2H_2$, $C_2H_4$ and other hydrocarbons and $H_2$ were removed from Ar, Xe up to a level of ~ $10^{-9}$ rel. vol. parts by $NiO/SiO_2$ – adsorbent at the temperature of ~ 450$^o$ C. The reaction $C_nH_m + NiO \rightarrow CO_2 + H_2O + Ni$ takes place in this adsorbent. The water was removed from all gases up to a level of ~ $10^{-4}$ torr by silica-gel. The oxygen was removed from $H_2O$ up to a level of ~ $10^{-13}$ rel. molar parts by $Ni/SiO_2$ – adsorbent at the temperature of ~ 200°C [13].

The purification systems for all these methods were constructed from stainless steel.

All vessels, pipes of the purification systems and chamber were heated and pumped out up to high vacuum before measurements.

In measurements three kinds of the signals shape were obtained (Fig. 2a,b,c). For all signals the quantity $\Delta t_1$ is the duration of ionization electrons collection from the particle track on the anode 5 of the chamber, i.e. this quantity is the duration of the signal caused by primary avalanches.

The quantity $\Delta t_2$ is the duration of the secondary avalanches. Point 1 in Fig. 2 a, b corresponds to the end of the primary avalanches.

The shape of the secondary signals is exponential. The total charge of the secondary avalanches is equal to ~ 50-70 % of the total charge in the proportional signal for the saturation region in dependence of admixture concentration.

The recorded signal for the mixture Ne + 30% $H_2$ (see Table I) is shown in Fig. 3.

From the analysis of the experimental data the following conclusions can be made:

1. The results obtained with $\alpha$ - source and $\beta$ - source do not qualitatively differ;
2. The secondary avalanches are produced in Penning mixtures – in our case Ar + $C_2H_2$, He + $CH_4$, Ne + $H_2$, Ar + Xe, Ne + Ar + $CH_4$ , $N_2$ + Ar + TMAE, Ar + $CH_4$);
3. The minimum admixture concentration with a low ionization potential, that gives secondary effects, may rich about 1 ppm (Ar + 1ppm $C_2H_2$);
4. The secondary avalanches take place in gases $CH_4$, $N_2$ + 10% $C_2H_2$ and $CF_4$ + 5% izo – $C_4H_{10}$ at a sacrifice of micro-admixtures of Ar (about of a several ppm) in commercial gases $CH_4$, $N_2$ and $CF_4$ which were purified in these experiments of $O_2$ – admixture only;
5. The mixture $CF_4$ + 5% izo – $C_4H_{10}$ is apparently Penning one because the excited molecules $CF_4$ are the intensive source of UV-photons [14];
6. The value of the total charge in the secondary processes depends on the concentration of admixture with a low ionization potential. For example, the total charge of secondary avalanches for the mixture Ar + 0.25 % $C_2H_2$ is equal to ~ 50% of the total charge in the proportional signal, but for the mixture Ar + 1ppm $C_2H_2$ – only 10%. The total charge of the secondary avalanches for the mixture Ne + 10% $H_2$ ($P_{abs}$ = 4 bar) is equal to ~ 66% as compared with 20% for the mixture Ne + 10 ppm $H_2$;
7. The duration of the secondary processes decreases with increasing of the electric field intensity in the proportional gap (Fig.4);
8. The duration of secondary processes does not practically depend on the drift electric field intensity;
9. The duration of secondary processes does not practically depend on the concentration of the admixture with a low ionization potential for the high values of concentration. For example, the quantity $\Delta t_2$ does not change in mixtures Ar + (10÷0.01)% $CH_4$. But the quantity $\Delta t_2$ for mixtures Ar + (1÷20) ppm $C_2H_2$ is equal to ~ 45μs as compared with the mixture Ar + 0.25% $C_2H_2$ for which $\Delta t_2$ = 100 μs. There is probably the saturation region in the dependence $\Delta t_2$ of admixture concentration. These dependences for different mixtures must be investigated in more detail;

10. Secondary Penning effects are absent in the one-component pure gases ($N_2$, Ne, $H_2$, He, Ar, $C_2H_2$) and in mixtures of pure molecular gases ($H_2$ + 0.5% $C_2H_2$, $H_2$ + 10% $CH_4$, $N_2$ + 33 ppm TMAE).

## II. Proportional discharge in the uniform electric field

The shape of the proportional avalanche discharge signal in the uniform electric field was measured with the chamber shown in Fig.5. The signals with parameters $\Delta t_1$ = 0.6 μs and $\Delta t_2$ = 10.5 μs (see Fig.6) were obtained for the chamber filling Ar + 1% $C_2H_2$ ($P_{abs}$ = 1 bar) and for voltages on electrodes $V_{an}$ = 500 V, $V_{cath}$ = -200 V. The shape of these signals is the same as in Fig.2a.

The signals with $\Delta t_1$ = 3 μs, $\Delta t_2$ = 25 μs were obtained for the filling 760 torr Ar + 4,5 torr $H_2O$ and for the voltages on electrodes $V_{an}$ = 500 V, $V_{cath}$ = -600 V.

It is necessary to note that coincidence of the signal shape in parallel-plate chamber with the shape of the electron component of the proportional discharge in chamber shown in Fig.1 gives an evidence for weak influence of the ion component of proportional discharge on the signal shape in the parallel-plate chamber. This result comes to an agreement with the results obtained in [15].

The signals with $\Delta t_1$ = 0.5 μs and without the secondary avalanches were obtained with the pure Ar-filling at the pressure of 1 bar and with voltages on the electrodes $V_{an}$ = 500 V, $V_{cath}$ = -200 V.

The authors [16] have observed secondary Penning avalanches in parallel-plate chamber for mixture Ar + 6 torr TEA, but they affirm, that secondary Penning avalanches are absent in mixture Ar + 10%$CH_4$ and there are in this mixture only the avalanches which are produced by photoelectrons from cathode of the chamber. Our measurements with chamber shown in Fig.1 show convincingly that there are secondary avalanches in mixtures Ar + (0.01 ÷ 10)% $CH_4$ at the sacrifice of Penning effects in the gas phase.

## III. Electroluminescent proportional chamber

In addition the proportional discharge was investigated with electroluminescent proportional chamber (Fig.7). The photomultiplier detected the photons from avalanches on the anode wires of the chamber. The signals with parameters $\Delta t_1$ = 0.3 μs, $\Delta t_2$ = 17 μs (Fig.8) were obtained by irradiation with the electrons of the chamber with a filling He + 10% $N_2$ + 100 ppm Xe ($P_{abs}$ = 1 bar) and with voltages on the electrodes $V_{an}$ = 5.2 kV, $V_{cath}$ = -220 V. This result confirms the existence of the secondary avalanches in Penning mixtures. The shape of these signals is the same as in Fig.2a.

The authors [17] have observed fast and slow component of the electro-luminescence in Ar. This result can be explained, if one suggests, that there was some admixture with a low ionization potential in Ar used for this experiment.

## IV. Discussing the experimental data

The observed secondary avalanches can only be explained by Penning effects because:

1. The secondary avalanches take place in Penning mixtures only,
2. The secondary effects are delayed relative to primary avalanches like the expected behaviour for Penning effects,
3. These effects can not be explained by photoelectrons from the cathode, because secondary avalanches are not observed for pure gases He, Ne, Ar.

From our results and the results of work [1] one can offer the following mechanism for the development of the secondary avalanches in Penning mixtures. Among other processes the excitation of the NG atoms X in metastable states $X_m^*$ takes place in the primary avalanches. The metastable Penning effect (2) takes place when atoms $X_m^*$ are colliding with the neutral molecules M of the admixture that have the ionization potential less than the excitation energy of the $X_m^*$ atoms.

The excitation of the NG atoms and molecules also occurs in avalanches with the following photons emitting [18] (processes (3), (4)). These photons ionize the molecules of impurities if their ionization potential less than the energy of photons (process (5)).

The electrons from processes (2) and (5) produce the secondary avalanches that in turn give the new excited states of NG atoms. As a result the avalanches are repeated many times during $\Delta t_2 \cong (1\div10)\times10$ μs. The delays in the appearance of the secondary avalanches are caused by the finite time of life of the metastable states and also by photons emitted in direction to the cathode.

At the moment of transition from the primary avalanches to the secondary avalanches a break in the shape of the signals takes place (point 1 in Fig.2a,b), which is caused by change in the intensity for the avalanches production. But it is necessary to note that it is possible to make the rise time of the primary signal equal to that of the secondary signal in point 1 by change of the drift electric field intensity.

The decrease of the secondary processes duration with the increase of the electric field intensity in the proportional gap (Fig.4) is caused by the acceleration of the entire process of proportional discharge development.

The increase of the volume of positive ion cloud at a sacrifice of the secondary avalanches and as a consequence the increase of its screening effect leads to a stop of the secondary avalanches, that is to a stop of the proportional discharge.

For one – component pure gases ($N_2$, Ne, $H_2$, Ar, He, $C_2H_2$ and so on) and for mixtures of pure molecular gases the duration of proportional signal is equal to the duration of primary avalanches. In this case the stop of the proportional discharge is caused by stopping of the arriving to the anode of ionization electrons from particle track.

In conclusion we compare our results with the results obtained in work [10], where the output of ionization electrons from α-tracks was measured in current conditions for the mixtures of He with Ar, Kr, $CO_2$, Xe, $N_2$, $C_2H_4$, $H_2$, of Ne with $H_2$, Ar, Xe and of Ar with $C_2H_2$ and $C_2H_4$ and it was shown that for concentration of admixtures $\geq 2\cdot 10^{-4}$ rel. vol. parts there is an increase of ionization electrons output up to 40% as compared with the pure He, Ne, Ar. Authors explain this additional ionization by production of metastable atoms with following process (1). The authors also note that micro-admixtures in commercial gases increase the ionization output, if these gases are used for chamber filling without addition of another gases. For removal of micro-admixtures influence the gases He, Ne, $H_2$ were purified by fine filtration at a temperature of 78 K.

Table I. Experimental data

| Gas | $P_{abs,}$ bar | RRad. Source | $V_{an}$, kV | $V_{cath}$, kV | Shape of the signal | | | $\Delta t_{1,}$ μs $\left(\frac{d\,(\Delta t_1)}{\Delta t_1}\cdot 10^2 \cong 10\%\right)$ | $\Delta t_{2,}$ μs $\left(\frac{d\,(\Delta t_2)}{\Delta t_2}\cdot 10^2 \cong 10\%\right)$ |
|---|---|---|---|---|---|---|---|---|---|
| | | | | | Fig. 2a | Fig.2b | Fig.2c | | |
| He + 3% $CH_4$ | 5 | β | 9 | –1 | 2a | | | 0.4 | 17 |
| He + 3% $CH_4$ | 17 | β | 15.5 | –1 | 2a | | | 1,0 | 20 |
| Ne + 30% $H_2$ | 1.5 | β | 4 | –1 | 2a | | | 1.1 | 10.8 |
| Ne + 30% Ar + 10% $CH_4$ | 1 | β | 4 | –2 | 2a | | | 1.8 | 12 |
| $CF_4$ + 5% izo-$C_4H_{10}$ | 2 | β | 16.8 | –2 | 2a | | | 0,5 | 39 |
| Ar +10% $H_2$ *⁾ | 1 | α | 3.7 | –2 | 2a | | | 2.5 | 20 |
| $CH_4$ | 1 | α | 5.3 | –1 | 2a | | | 1.5 | 17 |
| Ar + 10% $CH_4$ | 1 | α | 3.7 | –1 | 2a | | | 1.2 | 22.5 |
| Ar + 1% $CH_4$ | 1 | β | 4.8 | –1 | 2a | | | 1.1 | 20 |
| Ar + 0.1%$CH_4$ | 1 | β | 4.8 | –1 | 2a | | | 2.8 | 20 |
| Ar + 0.01%$CH_4$ | 1 | β | 4.5 | –1 | 2a | | | 5 | 20 |
| $N_2$ + 10% $C_2H_2$ | 1 | α | 6.5 | –1 | 2a | | | 2.5 | 18.5 |
| Ar + 0.25% $C_2H_2$ | 1 | α | 1.1 | –1 | 2a | | | 2.2 | 100 |
| Ar + 10% Xe**⁾ | 1 | α | 2.4 | –1 | 2a | | | 5 | 40 |
| $N_2$+10%Ar+33ppm TMAE | 1 | α | 8 | –1 | 2a | | | 4.5 | 22.5 |
| Ar + 20 ppm $C_2H_2$ | 1 | α | 1.75 | –1 | | 2b | | 12.8 | 48 |
| Ar + 1 ppm $C_2H_2$ | 1 | α | 2.6 | –1 | | 2b | | 12 | 40 |
| Ar | 1 | β | 4 | –2 | | | 2c | 13 | |
| Ne | 4 | β | 1 | –1 | | | 2c | ~175 | |
| He | 1 | β | 2.3 | –1 | | | 2c | 15 | |
| $H_2$ | 1 | β | 6.3 | –1 | | | 2c | 12 | |
| $C_2H_2$ | 0.2 | α | 3.5 | –2 | | | 2c | 10 | |
| $N_2$ | 1 | α | 8 | –0.6 | | | 2c | 12 | |
| $H_2$ + 0,5% $C_2H_2$ | 1 | α | 6 | –1 | | | 2c | 13 | |
| $H_2$ + 10% $CH_4$ | 1 | α | 5 | –1 | | | 2c | 14 | |
| $N_2$ + 33 ppm TMAE | 1 | α | 8 | –1 | | | 2c | 20 | |

*) In $H_2$ supplied by factory there were the micro-admixtures of hydrocarbons with a low ionization potential.

**) The mixture Ar + 10% Xe is a non metastable Penning mixture. In addition the secondary avalanches can be caused in this mixture by impurities of $Xe_2$ .

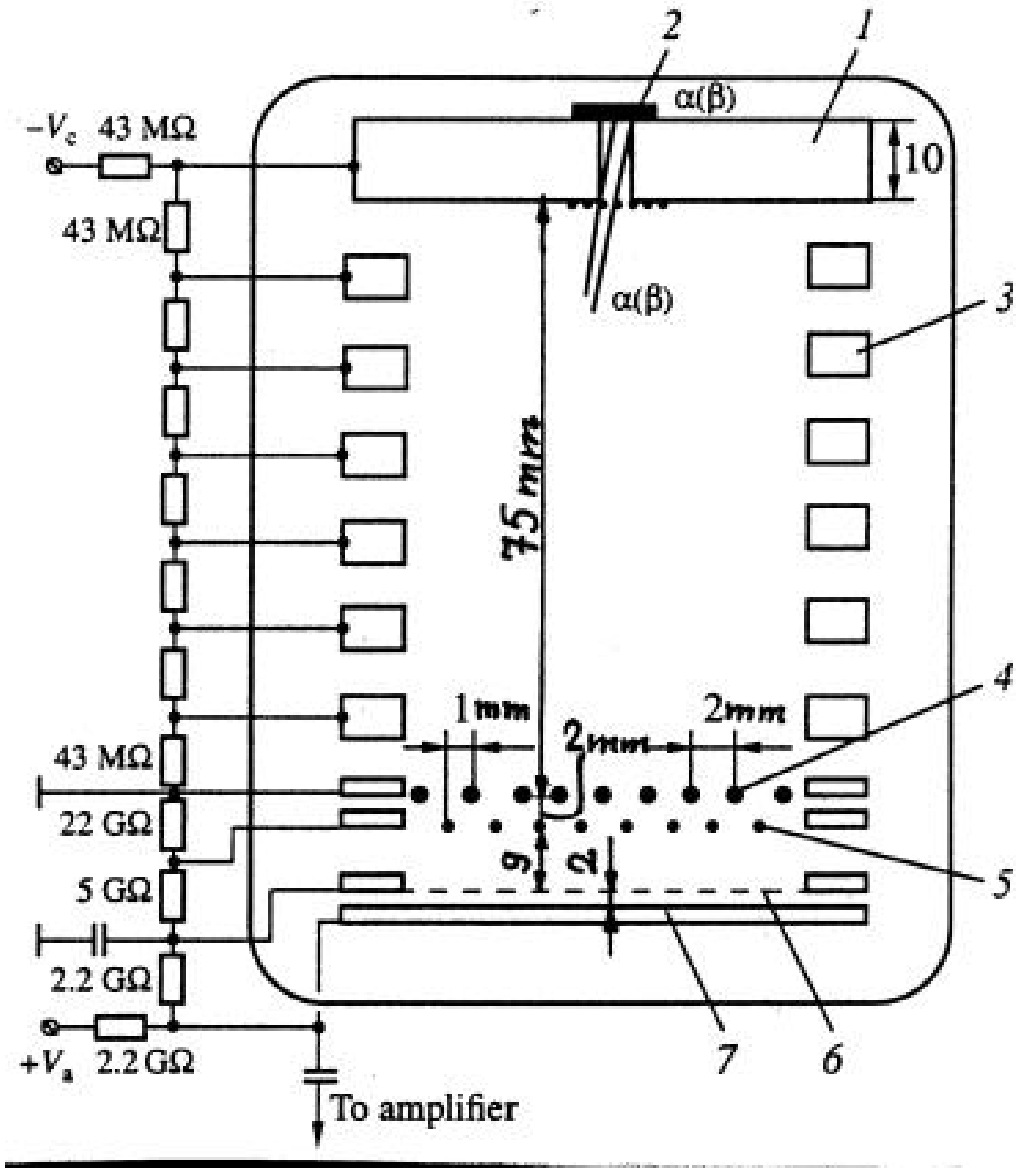

Fig.1. Ionization chamber with preliminary multiplication of ionization electrons and extraction of the electron component of the signal:

1 — cathode with collimator of 4 mm in diameter,

2 — α-source $Pu^{239}$ ($10^5 \alpha \cdot sec^{-1}$), or β-source $Ni^{63}$($10^4 \beta^- \cdot s^{-1}$),

3 — rings forming a drift uniform electric field,

4 — cathode of the proportional gap, $f_{wires}$ = 0.1 mm, pitch 2 mm,

5 — anode of the proportional gap, $f_{wires}$ = 20 μm, pitch 2 mm,

6 — cathode of the induction gap, $f_{wires}$ = 0.1 mm, pitch 1 mm,

7 — anode of the induction gap

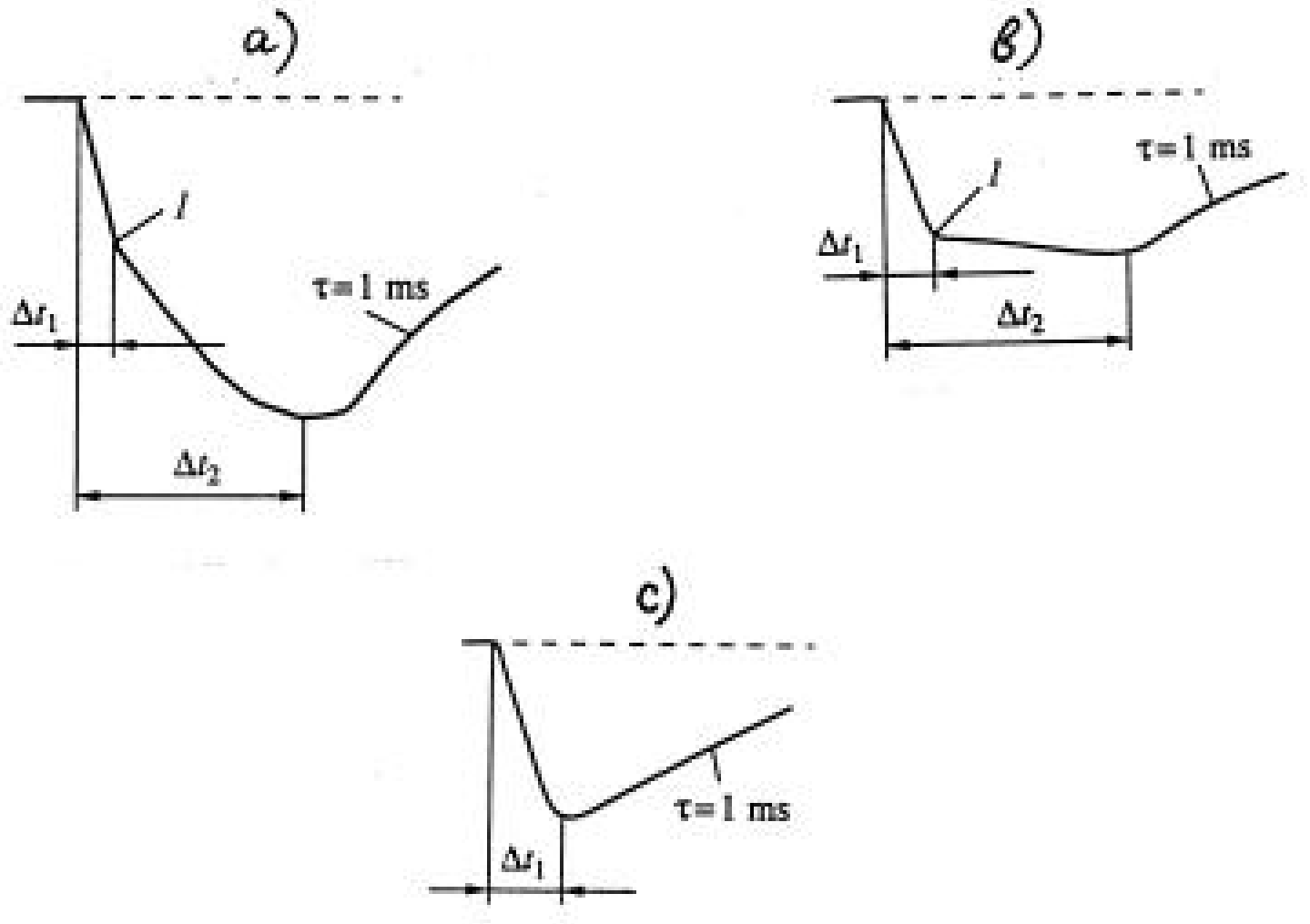

Fig.2. The shape of the signals obtained with the different gases:

a) with secondary avalanches (the break in point 1),

b) with secondary avalanches for low concentration of the admixture,

$\Delta t_1$ — the duration of the signal caused by primary avalanches,

$\Delta t_2$ — the duration of the signal caused by secondary avalanches

c) without secondary processes

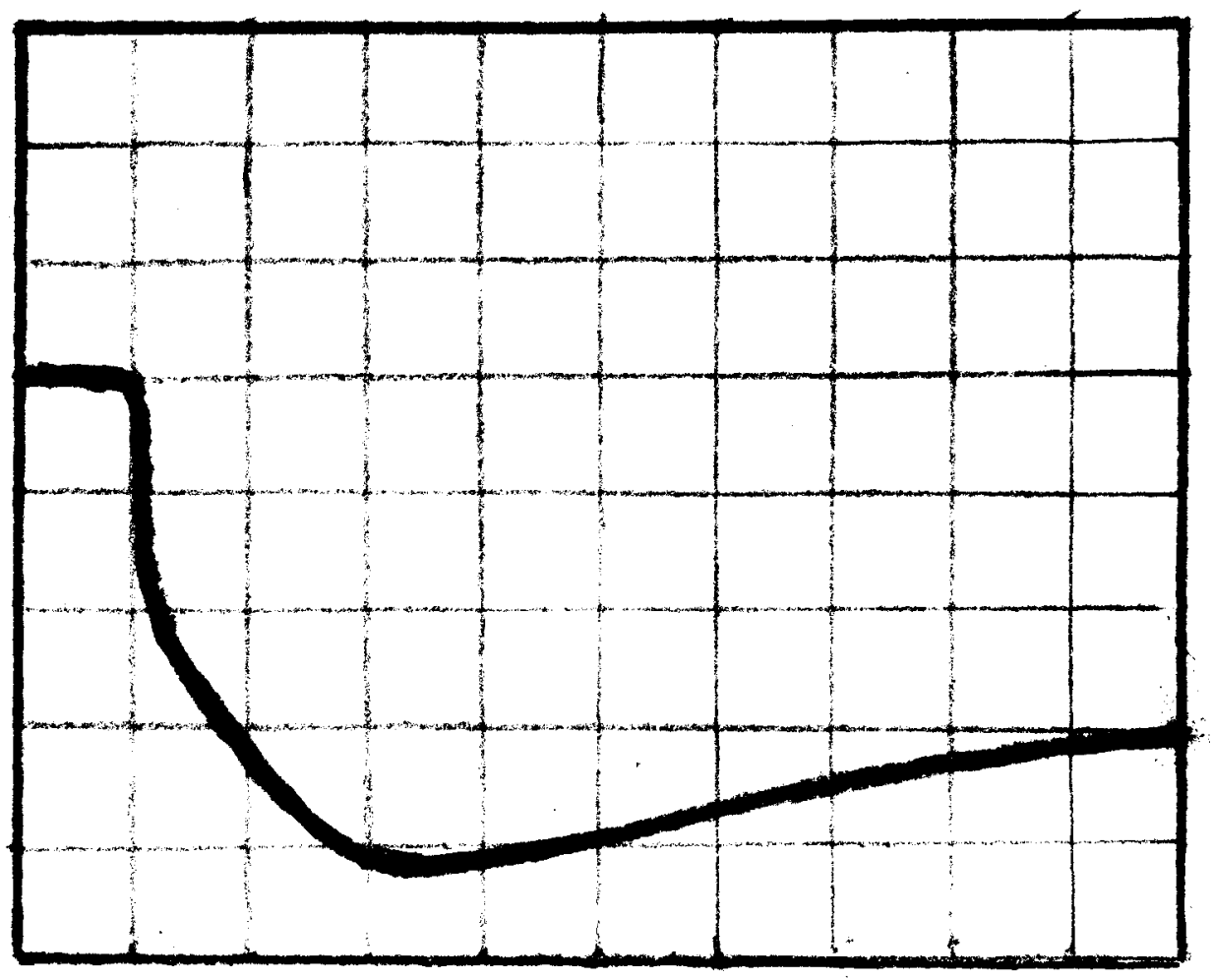

Fig. 3. The original recorded signal by the chamber shown in Fig. 1 for the mixture Ne + 30% $H_2$ (see Table I). One division on horizontal axes corresponds to 5 μs.

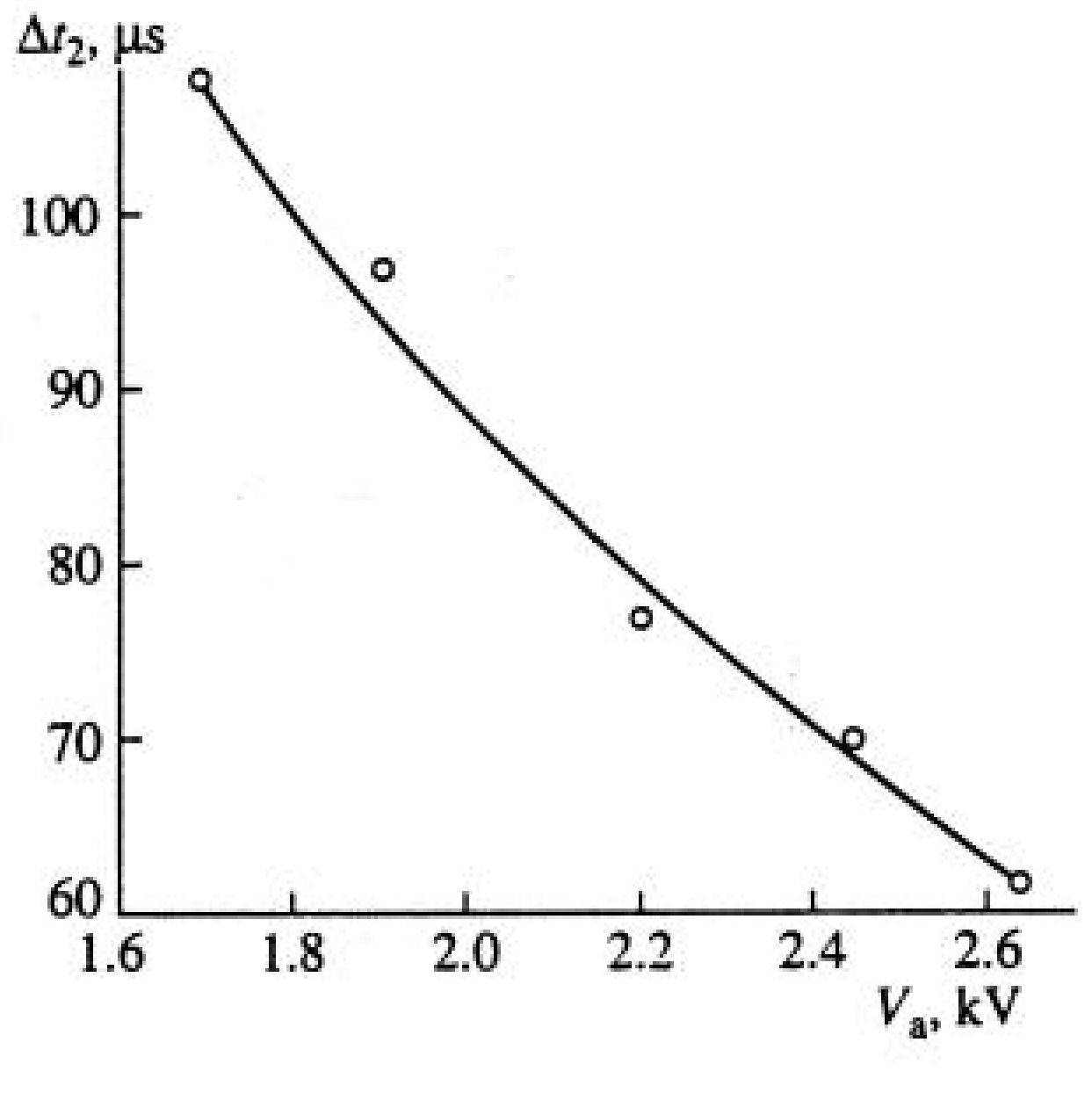

Fig.4. The dependence of $\Delta t_2$ from electric field intensity in proportional gap for the mixture Ar +0.5% $C_2H_2$ ($P_{abs}$ = 2 bar).

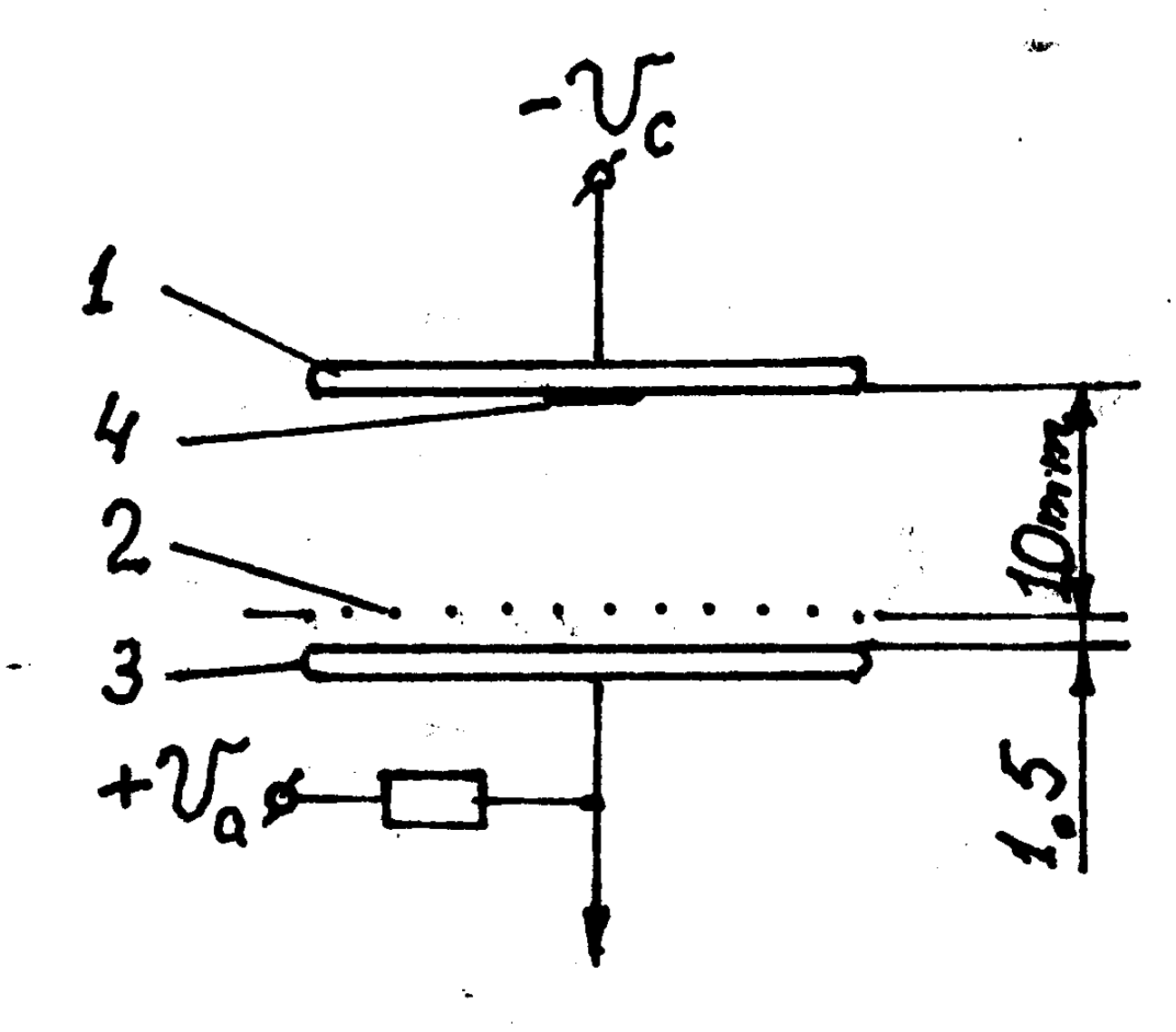

Fig.5. The parallel-plate avalanche chamber:

1 — cathode,

2 — screening greed with cells 0.5 x 0.5 mm, ϕ wires = 0.1 mm.

3 — anode.

4 — α - source $^{239}$Pu, $10^3 \alpha \cdot s^{-1}$

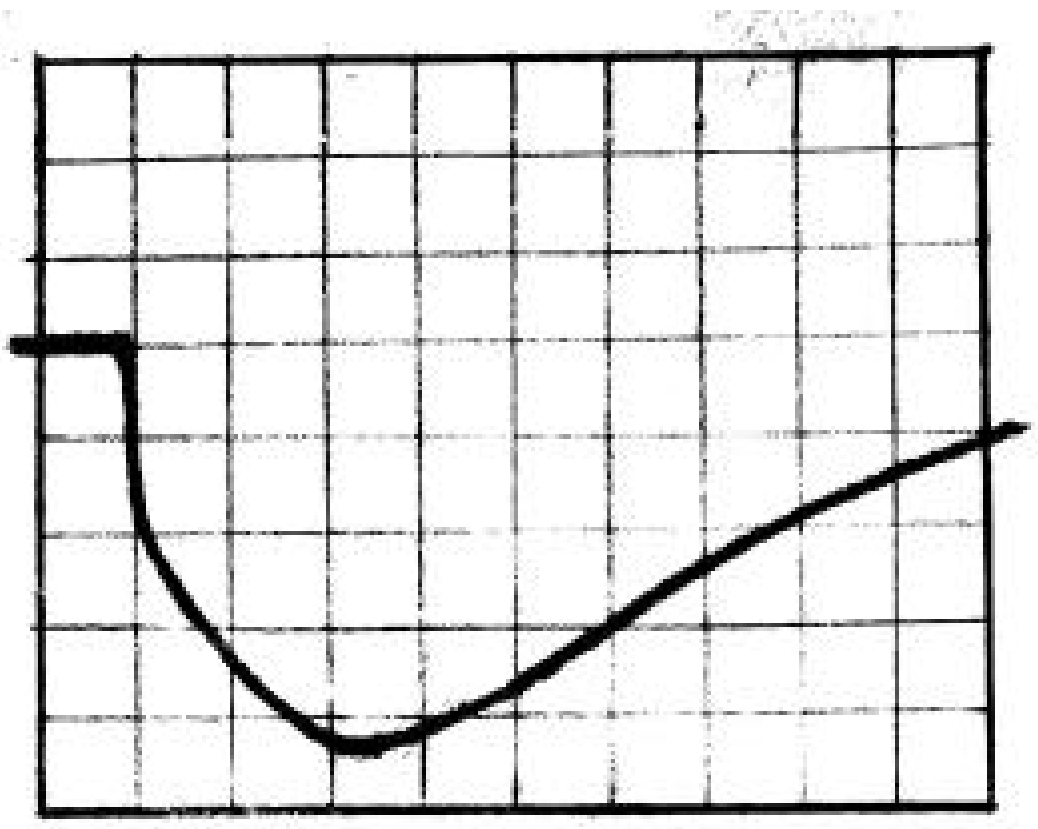

Fig.6. The original recorded signal by parallel – plate chamber shown in Fig.5 for the mixture Ar + 1% $C_2H_2$. One division on the horizontal axes corresponds to 5 μs.

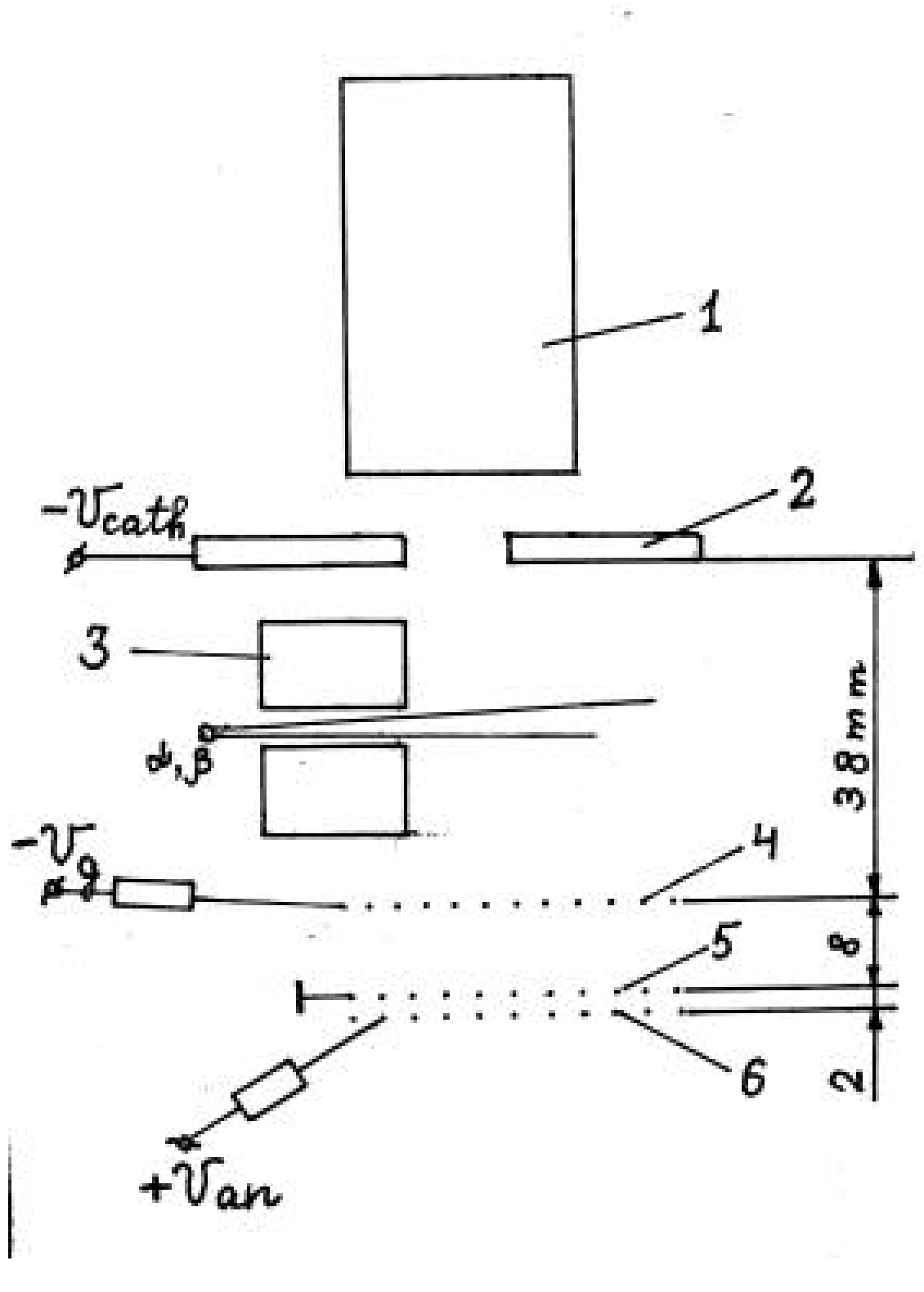

Fig.7. The electroluminescent proportional chamber:

1 — photomultiplier,

2 — cathode of the drift gap,

3 — collimator for α(β)-source,

4 — screening greed, ϕ wires = 0.1 mm, pitch 1 mm,

5 — cathode of the proportional gap, ϕ wires = 0.1 mm, pitch 1 mm,

6 — anode of the proportional gap, ϕ wires = 20 μm, pitch 1 mm.

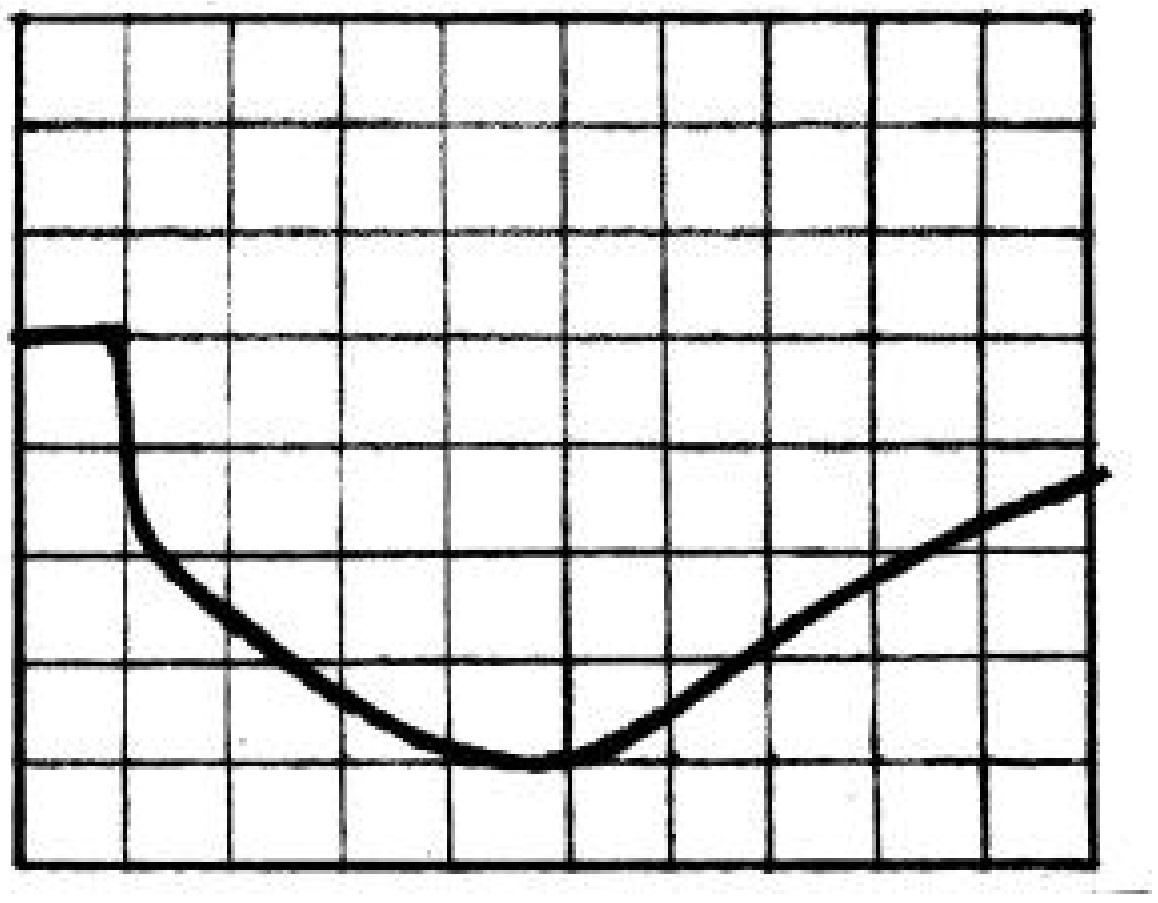

Fig.8. The original recorded signal by the electroluminescent chamber shown in Fig. 7 for the mixture He + 10% $N_2$ + 100 ppm Xe ($P_{abs}$ = 1 bar) and with voltages on the electrodes $V_{an}$ = 5.2 kV, $V_{cath}$ = -200 V. One division on the horizontal axes corresponds to 5 μs.